\documentclass[preprint,12pt]{elsarticle}
\usepackage{amsmath}
\usepackage{amssymb}
\usepackage{amsfonts}
\usepackage[usenames]{color}
\usepackage{dsfont}
\usepackage{epsfig}
\usepackage{graphicx}
\usepackage{hyperref}
\usepackage{relsize}
\usepackage{txfonts}
\usepackage{wasysym}
\newcommand{\dsty}[1]{\displaystyle{#1}}

\graphicspath{{./FIGURES/}}
\begin{document}
\begin{frontmatter}
\title{Addendum to ``On the consistency of MPS"}
\author[label1]{Antonio Souto-Iglesias\corref{cor1}}
\ead{antonio.souto@upm.es}
\author[label1]{Fabricio Maci\`{a}}
\ead{fabricio.macia@upm.es}
\author[label1]{Leo M. Gonz\'{a}lez}
\ead{leo.gonzalez@upm.es}
\author[label1]{Jose L. Cercos-Pita}
\ead{jl.cercos@upm.es}
\cortext[cor1]{Corresponding author.}
\address[label1]{Naval Architecture Department (ETSIN),\\Technical University of Madrid (UPM), 28040 Madrid, Spain.}
\begin{abstract}
The analogies between the Moving Particle Semi-implicit method (MPS) and
Incompressible Smoothed Particle Hydrodynamics method (ISPH) are
established in this note, as an extension of the MPS consistency
analysis conducted in \textit{Souto-Iglesias et al., Computer Physics Communications, 184(3), 2013.}
\end{abstract}
\end{frontmatter} 
\section{Introduction}
\label{s:intro}
Smoothed Particle Hydrodynamics (SPH) method
started in the seventies \cite{gingold1977} and was applied in the early nineties
to free-surface flows using an explicit approach with a weakly compressible
fluid model to numerically simulate liquid behavior \cite{mon1994b}.
In the mid nineties, the Moving Particle Semi-implicit method (MPS) appeared \cite{koshizuza1995,koshizuka1996}
imposing incompressibility with a projection scheme \cite{chorin_fractional_step_poisson_1968}.
Slightly later, 1999, a similar approach was followed
by Cummins and Rudman to obtain the first incompressible SPH (ISPH) model \cite{cummins1999,cummins2000}.
Although two MPS references were cited, no clear connections between ISPH and MPS were established.
A similar treatment was given to MPS in the posterior ISPH literature (e.g. \cite{lee2008,Rafiee20121,Xu20096703}).
In our opinion such MPS-SPH connections are clear after the equivalence between SPH and MPS approximation to first
and second order differential operators was established in \cite{soutoiglesias_etal_cpc2012_mps}.
Because these operators are the ones that play a major role in projection schemes,
this addendum to \cite{soutoiglesias_etal_cpc2012_mps} aims at clarifying the relationship between ISPH and
MPS methods. With this main goal set, this note is organized as follows: first the projection scheme
is reviewed, second, the MPS and ISPH implementations are
discussed and finally links between them are established.
%
%
\section{Projection fundamentals}
\label{sec:projection}
This section introduces the notation and reviews the fundamentals of the pressure
projection schemes.

In a projection, or fractional step, method \cite{chorin_fractional_step_poisson_1968}
for solving incompressible flows, the
pressure needed to enforce incompressibility is calculated by projecting
an estimate of the velocity field onto a divergence-free space.

The incompressible Navier-Stokes equations
in Lagrangian formalism are the field equations:
\begin{align}
\label{Navier_Stokes_inc1}
\frac{D \mathbf{r}}{D t} \, = &\, \mathbf{u}, \\
\label{Navier_Stokes_inc2}
\dsty{\nabla\cdot}\mathbf{u}  \, =& \, 0, \\
\dsty{\frac{D \mathbf{u}}{D t}} \, = &\, \mathbf{g} \, + \, \frac{\nabla \cdot \varmathbb{T}}{\rho}.
\label{Navier_Stokes_inc}
\end{align}
%
%
%
%
where $\rho$ stands for the fluid density  and $\mathbf{g}$ is a generic
external volumetric force field. The flow velocity $\mathbf{u}$ is defined as the material
derivative of a fluid particle with position $\mathbf{r}$.
$\varmathbb{T}$ denotes the stress tensor of a Newtonian incompressible fluid:
\begin{equation}
\label{Tstress}
\varmathbb{T}  \, =  - P \,\boldsymbol{\mathit{I}}\, + \, 2 \, \mu \, \varmathbb{D} \,,
\end{equation}
in which $P$ is the pressure, $\varmathbb{D}$ is the rate of deformation tensor
$(\varmathbb{D} = ( \nabla \mathbf{u} + \nabla \mathbf{u}^{T} )/2)$
and $\mu$ is the dynamic viscosity.
With this notation, the divergence of the stress tensor $\varmathbb{T}$
is computed as:
\begin{equation}
\label{varmathbbT}
\nabla \cdot \varmathbb{T} \, = \, - \, \nabla P \,  + \, \mu \nabla^2 \mathbf{u} \,.
\end{equation}
In order to numerically integrate these equations, the fluid domain is discretized in a set of particles
whose positions are $\mathbf{r}_a$. For the fractional step method, in a generic time step $n$,
first, the particle positions are advected with the available velocity, $\mathbf{u}_a^n$, considering
a time step $\Delta t$:
\begin{equation}
\label{rstar}
\mathbf{r}_a^*=\mathbf{r}_a^n+\Delta t \left(\mathbf{u}_a^n \right).
\end{equation}
Second, considering the advected positions to evaluate the viscous interactions,
an intermediate velocity field $\mathbf{u^*}$ is explicitly computed using
the momentum equation but ignoring the pressure term:
\begin{equation}
\label{ustar}
\mathbf{u}_a^*=\mathbf{u}_a^n+\Delta t
\left(
  \mu \nabla^2 \mathbf{u}
\right)_a^* + \Delta t \, \mathbf{g}.
\end{equation}
Third, the zero divergence condition
is imposed on the velocity field at
the next time step, thus obtaining the Poisson equation for the pressure:
\begin{equation}
\label{eq:poissonMPS}
\left(\nabla^2 P\right)_a^{n+1}=\frac{\rho}{\Delta t}\left(\nabla\cdot\mathbf{u}_a^*\right).
\end{equation}
Once the pressure is found, pressure gradients are
computed, the velocity is updated:
\begin{equation}
\label{unplus1}
\mathbf{u}_a^{n+1}=\mathbf{u}_a^*-\Delta t
\frac{1}{\rho}
\left(
  \nabla P\right
)_a^{n+1},
\end{equation}
and the particle positions are modified, usually with an implicit scheme:
\begin{equation}
\label{rnplus1euler}
\mathbf{r}_a^{n+1}=\mathbf{r}_a^n+\Delta t \left(\mathbf{u}_a^{n+1} \right),
\end{equation}
or a Crank-Nicholson one, yielding:
\begin{equation}
\label{rnplus1crank}
\mathbf{r}_a^{n+1}=\mathbf{r}_a^n+\Delta t
\left(
\frac{\mathbf{u}_a^{n+1}+\mathbf{u}_a^{n}}{2}
\right).
\end{equation}
Let us see how this scheme is applied in MPS and in ISPH. 
\section{Moving Particle Semi-implicit method (MPS)}
Let us focus on the MPS time integration scheme \cite{youngyoon1999}, in which
the Poisson problem for the pressure is written as:
\begin{align}
\label{eq:poissonmpslapl}
\langle
  \nabla^2 P^{n+1}
\rangle_a^{MPS}
&
=
\frac{\rho}{\Delta t}
\langle
  \nabla\cdot\mathbf{u^*}
\rangle_a^{MPS},  \\
\label{eq:poissonmpsadvc}
\frac{\mathbf{u}_a^{n+1}-\mathbf{u}_a^{*}}{\Delta t}
&
=
-\frac{1}{\rho}
\langle
  \nabla P^{n+1}
\rangle_a^{MPS}.
\end{align}
%
%
The problem setup therefore respects the formalism described in section \ref{sec:projection},
but referred to the smoothed operators. The positions are in the MPS literature mostly
advected with the first order implicit scheme (\ref{rnplus1euler}).

%
If the operators \ref{eq:poissonmpslapl}-\ref{eq:poissonmpsadvc} are written in their MPS integral form \cite{soutoiglesias_etal_cpc2012_mps},
equations (\ref{eq:poissonmpslapl}), (\ref{eq:poissonmpsadvc}) become:
\begin{align}
\frac{2 A_0}{\left(r_e\right)^{2} A_2}
\int_{\mathbb{R}^d}
    \left[
        P(\mathbf{x}^{\prime}) - P(\mathbf{x}_a)
    \right]
    w
    \left(
            \frac
            {\left|\mathbf{x}_a-\mathbf{x^{\prime}}\right|}
            {r_e}
    \right)
    d\mathbf{x^{\prime}}
& = \nonumber \\
\label{eq:poissonmps1}
\frac{\rho}{\Delta t}
\int_{\mathbb{R}^d}
&
    \frac
    {
        \left(
            \mathbf{u^\prime} - \mathbf{u}_a
        \right)
        \cdot
        \left(
            \mathbf{x^\prime} - \mathbf{x}_a
        \right)
    }
    {
        {\left|\mathbf{x^{\prime}}-\mathbf{x}_a\right|^2}
    }
    w
    \left(
            \frac
            {\left|\mathbf{x}_a-\mathbf{x^{\prime}}\right|}
            {r_e}
    \right)
    d\mathbf{x^{\prime}},
\end{align}
\begin{equation}
\label{eq:poissonmps2}
\frac{\mathbf{u}_a^{n+1}-\mathbf{u}_a^{*}}{\Delta t}
=
-\frac{d}{\rho \left(r_e\right)^d A_0 }
\int_{\mathbb{R}^d}
    \frac
    {
        P(\mathbf{x^{\prime}}) - P(\mathbf{x}_a)
    }
    {
        {\left|\mathbf{x^{\prime}}-\mathbf{x}\right|^2}
    }
    \left(\mathbf{x^{\prime}}-\mathbf{x}_a\right)
    w
    \left(
            \frac
            {\left|\mathbf{x}_a-\mathbf{x^{\prime}}\right|}
            {r_e}
    \right)
    d\mathbf{x^{\prime}},
\end{equation}
where $d$ is the dimensionality of the problem, $w$ is the MPS weighting function, $r_e$ is the cut-off radius of $w$ and $A_0$, $A_2$ are constants
which depend on the specific form of the weighting function \cite{soutoiglesias_etal_cpc2012_mps}.

\section{Incompressible SPH (ISPH)}
The system solved in ISPH is the same as in MPS \cite{lee2008}:
\begin{align}
\langle
  \nabla^2 P^{n+1}
\rangle_a^{SPH}
&
=
\frac{\rho}{\Delta t}
\langle
  \nabla\cdot\mathbf{u^*}
\rangle_a^{SPH}, \\
\frac{\mathbf{u}_a^{n+1}-\mathbf{u}_a^{*}}{\Delta t}
&
=
-\frac{1}{\rho}
\langle
  \nabla P^{n+1}
\rangle_a^{SPH},
\end{align}
%
%
%
where the previously mentioned operators can be written in the integral SPH formalism according to the consistency
analysis of \cite{espanol2003,mon2005,MaciaetalPTP}.
\begin{align}
\label{eq:poissonsph1}
-\frac{2}{h^{d+1}}
\int_{\mathbb{R}^{d}}
\frac{P\left(  \mathbf{x}^{\prime
}\right)  -P\left(  \mathbf{x}_a\right)  }{\left\vert \mathbf{x}^{\prime
}-\mathbf{x}\right\vert }\tilde{W}^{\prime}\left(  \frac{\left\vert
\mathbf{\mathbf{x}^{\prime}-\mathbf{x}}\right\vert }{h}\right)  d\mathbf{x}%
^{\prime}
&
=\nonumber \\
\frac{\rho}{\Delta t}
\frac{1}{h^{d+1}}\int_{\mathbb{R}^{d}}
&
\frac{
        \left(
            \mathbf{u}^{\prime}
            -
            \mathbf{u}_a
        \right)
        \cdot
        \left(
            \mathbf{x} - \mathbf{x^\prime}
        \right)
}
{
    \left\vert \mathbf{x}-\mathbf{x}^{\prime}\right\vert
}
\tilde{W}^{\prime}
\left(  \frac{\left\vert \mathbf{\mathbf{x-x}%
}^{\prime}\right\vert }{h}\right)
d\mathbf{x}^{\prime},
\end{align}
\begin{equation}
\label{eq:poissonsph2}
\frac{\mathbf{u}_a^{n+1}-\mathbf{u}_a^{*}}{\Delta t}
=
-\frac{1}{\rho}
\frac{1}{h^{d+1}}\int_{\mathbb{R}^{d}}
\frac
{
    P
    \left(  \mathbf{x}^{\prime}\right)
    -
    P
    \left(  \mathbf{x}_a\right)
}
{
    \left\vert \mathbf{x}_a-\mathbf{x}^{\prime}\right\vert
}
\left(
  \mathbf{x}_a-\mathbf{x}^{\prime}
\right)
\tilde{W}^{\prime}
\left(
\frac
{
  \left\vert
  \mathbf{x}^{\prime}-\mathbf{x}_a
  \right\vert
}
{h}
\right)
d\mathbf{x}^{\prime},
\end{equation}
where $\tilde{W}:\mathbb{R\rightarrow R}$ is a nonnegative
differentiable function such that:%
\begin{equation}
\int_{\mathbb{R}^{d}}
\tilde{W}\left(
\left\vert \mathbf{x}\right\vert
\right)
d\mathbf{x}=1.
\end{equation}
and the SPH kernel $W$ is defined on the basis of  $\tilde{W}$ as:
\begin{equation}
\label{eq:sphkernel}
W\left(  \mathbf{x};h\right)  =\frac{1}{h^{d}}\tilde{W}\left(  \left\vert \frac{\mathbf{x}}{h}\right\vert \right)  \text{.}%
\end{equation}
with $h$ being proportional to the cut-off radius of the kernel.

Although different options are available for the gradient formula, such as Monaghan's
symmetrized one \cite{mon1994b} commonly used in SPH, their order is $h^2$ regardless of the formula.
%
%
%

\section{Passing from MPS to ISPH and the other way around}
Equations (\ref{eq:poissonmps1})-(\ref{eq:poissonmps2}) and
(\ref{eq:poissonsph1})-(\ref{eq:poissonsph2}) corresponding to MPS and ISPH respectively,
present at the formal level some structural similarities.
Hence it is only natural to explore deeper relations between these two methods.

Using the results of Souto-Iglesias et al.\cite{soutoiglesias_etal_cpc2012_mps} to establish the MPS
consistency, it is possible to obtain equivalent operators in SPH from MPS ones and viceversa.
As a matter of fact, these equivalences at the integral level are based on formulae
that relate the MPS weighting function with the SPH kernel.

More precisely, the MPS scheme (\ref{eq:poissonmps1})-(\ref{eq:poissonmps2}) can be
written using the ISPH formalism as:
\begin{align}
\label{eq:poissonsph1MPS}
-\frac{2}{h^{d+1}}
\int_{\mathbb{R}^{d}}
\frac{P\left(  \mathbf{x}^{\prime
}\right)  -P\left(  \mathbf{x}_a\right)  }{\left\vert \mathbf{x}^{\prime
}-\mathbf{x}\right\vert }\tilde{W}_\Delta^{\prime}\left(  \frac{\left\vert
\mathbf{\mathbf{x}^{\prime}-\mathbf{x}}\right\vert }{h}\right)  d\mathbf{x}%
^{\prime}
&
= \nonumber \\
\frac{\rho}{\Delta t}
\frac{1}{h^{d+1}}\int_{\mathbb{R}^{d}}
&
\frac{
        \left(
            \mathbf{u}^{\prime}
            -
            \mathbf{u}_a
        \right)
        \cdot
        \left(
            \mathbf{x} - \mathbf{x^\prime}
        \right)
}
{
    \left\vert \mathbf{x}-\mathbf{x}^{\prime}\right\vert
}
\tilde{W}_\nabla^{\prime}
\left(  \frac{\left\vert \mathbf{\mathbf{x-x}%
}^{\prime}\right\vert }{h}\right)
d\mathbf{x}^{\prime},
\end{align}
\begin{equation}
\label{eq:poissonsph2MPS}
\frac{\mathbf{u}_a^{n+1}-\mathbf{u}_a^{*}}{\Delta t}
=
-\frac{1}{\rho}
\frac{1}{h^{d+1}}\int_{\mathbb{R}^{d}}
\frac
{
    P
    \left(  \mathbf{x}^{\prime}\right)
    -
    P
    \left(  \mathbf{x}_a\right)
}
{
    \left\vert \mathbf{x}_a-\mathbf{x}^{\prime}\right\vert
}
\left(
  \mathbf{x}_a-\mathbf{x}^{\prime}
\right)
\tilde{W}_\nabla^{\prime}
\left(
\frac
{
  \left\vert
  \mathbf{x}^{\prime}-\mathbf{x}_a
  \right\vert
}
{h}
\right)
d\mathbf{x}^{\prime},
\end{equation}
using different kernels for first and second order differential operators:
\begin{align}
\label{Wwgradequiv}
\tilde{W}_\nabla(q)
&
=
-\frac{d}{A_0}
\int_0^q \frac{1}{s}w(s) ds + C_1,\\
\label{Wwlaplequiv}
\tilde{W}_\Delta(q)
&
=
-\frac{d}{A_2}
\int_0^q {s}\,w(s) ds + C_2.
\end{align}
with $q=|\mathbf{x}/h|$. Constants $C_1$, $C_2$ are obtained by imposing $\tilde{W}_\nabla(1)=\tilde{W}_\Delta(1)=0$ \cite{soutoiglesias_etal_cpc2012_mps}.

%

On the other hand, the ISPH integral formulation of the problem as expressed in equations
(\ref{eq:poissonsph1})-(\ref{eq:poissonsph2}) can be seen from the MPS point of view provided
different weighting functions are used to approximate first and second order differential operators, respectively:
\begin{align}
\label{eq:poissonmps1SPH}
\frac{2 A_0}{\left(r_e\right)^{2} A_2}
\int_{\mathbb{R}^d}
    \left[
        P(\mathbf{x}^{\prime}) - P(\mathbf{x}_a)
    \right]
    w^\Delta
    \left(
            \frac
            {\left|\mathbf{x}_a-\mathbf{x^{\prime}}\right|}
            {r_e}
    \right)
    d\mathbf{x^{\prime}}
&
= \nonumber \\
\frac{\rho}{\Delta t}
\int_{\mathbb{R}^d}
&
    \frac
    {
        \left(
            \mathbf{u^\prime} - \mathbf{u}_a
        \right)
        \cdot
        \left(
            \mathbf{x^\prime} - \mathbf{x}_a
        \right)
    }
    {
        {\left|\mathbf{x^{\prime}}-\mathbf{x}_a\right|^2}
    }
    w^\nabla
    \left(
            \frac
            {\left|\mathbf{x}_a-\mathbf{x^{\prime}}\right|}
            {r_e}
    \right)
    d\mathbf{x^{\prime}}
\end{align}
\begin{equation}
\label{eq:poissonmps2SPH}
\frac{\mathbf{u}_a^{n+1}-\mathbf{u}_a^{*}}{\Delta t}
=
-\frac{d}{\rho \left(r_e\right)^d A_0 }
\int_{\mathbb{R}^d}
    \frac
    {
        P(\mathbf{x^{\prime}}) - P(\mathbf{x}_a)
    }
    {
        {\left|\mathbf{x^{\prime}}-\mathbf{x}\right|^2}
    }
    \left(\mathbf{x^{\prime}}-\mathbf{x}_a\right)
    w^\nabla
    \left(
            \frac
            {\left|\mathbf{x}_a-\mathbf{x^{\prime}}\right|}
            {r_e}
    \right)
    d\mathbf{x^{\prime}},
\end{equation}
with
\begin{align}
\label{eq:wW_wendland}
w^\nabla(q)&:=- \frac{q}{d} \tilde{W}^{\prime}(q),\\
\label{eq:Ww_wendland}
w^\Delta(q)&:=- \frac{d}{q} \tilde{W}^{\prime}(q).
\end{align}
The equivalences established so far do not depend neither on the time integration scheme used nor on
the implementation of  boundary conditions. Most importantly, these equivalences
are not affected by the discretization of the smoothed operators,
where mass-carrying particles are used to represent
the integrals in both methods \cite{mon1992,koshizuka1996}.
Summarizing, any MPS formulation can be equivalently reformulated as an ISPH scheme and reciprocally.

At this point, it becomes clear that the question of comparing
the solution obtained through an MPS based method to one obtained by an ISPH implementation
reduces to that of understanding the sensitivity of MPS to the weighting function used (or equivalently that of ISPH to
the kernel considered). This remark is relevant since the choice of the kernel may have a significant influence on
several properties of the numerical scheme, namely:
stability (see e.g. \cite{Swegle+etal:1995,dehnen_aly_wendland_2012,colagrossi_etal_pre2013_standingwave}), accuracy \cite{Quinlan_06,Amicarelli2011279} and thermodynamic consistency \cite{violeau_2009}.

\section{Alternative RHS formulation}
\subsection{General}
The corrective term on the right hand side of the Poisson equation (\ref{eq:poissonmpslapl})
can be reformulated using the continuity equation to estimate the divergence of the velocity field:
\begin{equation}
  \nabla\cdot\mathbf{u^*}=-\frac{1}{\rho}\frac{d\rho}{dt}.
\end{equation}
This leads to an alternative way to write  equation (\ref{eq:poissonmpslapl}), namely:
\begin{align}
\label{eq:poissonmpslapl2}
  \nabla^2 P^{n+1}
&
=
-\frac{\rho_0}{\Delta t^2}\frac{\rho^*-\rho_0}{\rho_0},
\end{align}
in which $\rho_0$ is the reference density and $\rho^*$ is the intermediate time step, which is obtained at the discrete level by summations across neighboring particles either using a MPS weight function or an SPH kernel. These summations may reflect an excess or defect of local mass, a consequence of the fact that the intermediate velocity field $\mathbf{u^*}$ may not satisfy the divergence free constraint.
\subsection{MPS approximation}
\label{ss:mpssigma}
This alternative formulation has been used in a large proportion of the MPS literature \cite{ShaoLo_2003,Tsukamoto_cheng_2011,Khayyer2010_aor} including the seminal papers by Koshizuka and collaborators \cite{koshizuza1995,koshizuka1996}.

The intermediate density $\rho^*$ is obtained in MPS for each particle $a$ as \cite{youngyoon1999}:
\begin{equation}
\label{eq:rhoastar}
\langle \rho^*\rangle_a^{MPS}
=
\frac
{
    m \langle n \rangle_a
}
{
\int_{\mathbb{R}^d}
w
\left(
        \frac
        {\left|\mathbf{x}\right|}
        {r_e}
\right)
\,d\mathbf{x}
}
=
\frac
{
    m \sum_{b\in J_a}
w
\left(
        \frac
        {\left|\mathbf{x}_a-\mathbf{x}_b\right|}
        {r_e}
\right)
}
{
A_0 r_e^d
}
.
\end{equation}
In this formula $m$ is the mass of each individual particle and $\langle n \rangle_a$ is a particle number density defined as:
\begin{equation}
\label{eq:ni}
\langle n \rangle_a :=
\sum_{b\in J_a}
w
\left(
        \frac
        {\left|\mathbf{x}_a-\mathbf{x}_b\right|}
        {r_e}
\right),
\end{equation}
where $J_a$ is the set of indexes corresponding to neighboring particles. When $w$ is singular for argument zero (e.g. \cite{koshizuka1996,Khayyer2010_aor}) this index set does not include the particle $a$ itself .

Since the MPS weight function $w$ is positive, isotropic and with compact support, an
SPH kernel can be constructed from $w$ as:
\begin{equation}
\label{eq:sphkernel}
W\left(  \mathbf{x};r_e\right)  =\frac{1}{A_0 r_e^{d}}w\left(  \left\vert \frac{\mathbf{x}}{r_e}\right\vert \right)  \text{,}%
\end{equation}
It is straightforward to see that the volume integral of this function equals one, a necessary condition for $W$ to be
a well defined kernel. Let us denote this kernel as $W_\Sigma$.
%

Considering this new kernel, one can write:
\begin{equation}
\label{eq:rhoastar3}
\langle \rho^*\rangle_a^{MPS}
=
m \sum_{b\in J_a} W_\Sigma\left(  \mathbf{x}_b-\mathbf{x}_a;r_e\right)=\langle \rho^*\rangle_a^{SPH},
\end{equation}
and this summation becomes the canonic SPH approximation to the local value of the density,
$\langle \rho^*\rangle_a^{SPH}$ \cite{mon2005}, where, as in \cite{soutoiglesias_etal_cpc2012_mps},
the cut-off radius $r_e$ is identified to the SPH smoothing length.

Therefore, for each MPS weight function $w$ a well defined SPH kernel $W_\Sigma$ exists, providing an equivalent local estimation
of the density field. Note that in the cases when $w$ is singular at the origin, the SPH kernel $W_\Sigma$ cannot be used to approximate differential operators.
\subsection{ISPH approximation}
The idea of using a corrective term based on density variations with respect to the reference density can
also be found in the ISPH literature \cite{ShaoLo_2003,shao}, although, originally, in the works of Cummins and Rudman \cite{cummins1999}, such a term was based on the velocity divergence.
We should also mention Zhou \cite{zhou_2010_phd} who even used a mixed formulation,
computing the Poisson equation RHS by weight averaging the velocity divergence and the density correction terms.

Analogously to section \ref{ss:mpssigma}, from equations (\ref{eq:sphkernel}-\ref{eq:rhoastar3})
it follows that given an SPH kernel it is possible to find
an infinite number of MPS weight functions which provide the same local estimation of the density field. However, all these
MPS weight functions are proportional.
\subsection{Summary}
If the corrective source term is based on the density variation, establishing the equivalence between SPH and MPS requires defining a new SPH kernel from the MPS weight function. This kernel adds to the ones that are necessary for MPS operators to consistently represent first and second order differential operators \cite{soutoiglesias_etal_cpc2012_mps}. Therefore, three SPH kernels need to be defined from each MPS function in order to pass from MPS to SPH:
\begin{align}
\tilde{W}_\Sigma\left(q\right)
&
=
\frac{1}{A_0}w\left(q\right)  \text{,}
\nonumber
\\
\tilde{W}_\nabla(q)
&
=
-\frac{d}{A_0}
\int_0^q \frac{1}{s}w(s) ds + C_1,
\nonumber
\\
%
\tilde{W}_\Delta(q)
&
=
-\frac{d}{A_2}
\int_0^q {s}\,w(s) ds + C_2.
\nonumber
\end{align}
Each kernel is re-scaled by introducing the cut-off radius:
\begin{equation}
\label{eq:sphmpsgrad}
W_\Box\left(  \mathbf{x};r_e\right)  =\frac{1}{r_e^{d}}\tilde{W}_\Box\left(  \left\vert
\frac{\mathbf{x}}{r_e}\right\vert \right).%
\end{equation}
It is also possible to build equivalent MPS weight functions from a given SPH kernel. For details we refer the reader to \cite{soutoiglesias_etal_cpc2012_mps}.

A final outcome of the present analysis is to provide MPS with a consistent interpolation formula for any flow field which allows to compute it at any point in space regardless of whether a particle exists there:
\begin{equation}
\label{eq:fmps}
\langle f \rangle_a^{MPS}
=
 \sum_{b\in J_a} \frac{m}{\rho_b}\, f_b\, W_\Sigma\left(  \mathbf{x}_b-\mathbf{x}_a;r_e\right).
\end{equation}
Due to its equivalence with SPH, the order of this formula is $\mathcal{O}(r_e^2)$.
%

\section{Conclusions}
Analogies between the Moving Particle Semi-implicit method (MPS) and
Incompressible Smoothed Particle Hydrodynamics method (ISPH) have been
discussed in the present paper, showing that any MPS scheme can be reformulated
as an ISPH one and viceversa.

This equivalence is based on reformulating the MPS density interpolation formula
and the first and second order differential operators within the ISPH framework by defining different
SPH kernels for each of these operators.

The numerical analysis of meshless methods presents unsettled issues concerning stability,
conservation properties, computational efficiency, etc.. that are still unresolved today.
We think that the present note could be useful in providing the framework needed to
view the significant amount of work on SPH and MPS on these topics from
a new perspective and help the progress of meshless methods.
\section*{Acknowledgments}
The research leading to these results has received funding from the
Spanish Ministry for Science and Innovation under
grant TRA2010-16988 ``\textit{Ca\-rac\-te\-ri\-za\-ci\'on Num\'erica y
Experimental de las Cargas Fluido-Din\'amicas en el transporte de Gas Licuado}'' .
The authors thank Hugo Gee for the English proof-reading of the manuscript.
%

\bibliographystyle{model1a-num-names}

\end{document}